\documentclass[preprint,showpacs,epsfig]{revtex4}
\usepackage{graphicx}

\begin{document}

\draft
\title{Resonances in a circular dielectric cavity}
\author{Jung-Wan Ryu$^{1}$}
\email[E-mail address:]{jungwanryu@pnu.ac.kr}
\author{Sunghwan Rim$^2$}
\author{Young-Jai Park$^2$}
\author{Chil-Min Kim$^2$}
\author{Soo-Young Lee$^{3}$}
\email[Corresponding author. Tel:+82-2-880-1468, Fax: +82-2-876-3973, E-mail address: ]
{pmzsyl@naver.com}
\affiliation{$^1$ Department of physics, Pusan National University,
Busan, 609-735, Korea}
\affiliation{$^2$ Department of Physics, Sogang University,
Seoul 121-742, Korea}
\affiliation{$^3$ School of Physics and Astronomy, Seoul National University,
Seoul, 151-742, Korea}

\begin{abstract}
We study resonance distributions in a circular dielectric cavity.
It is shown that the decay-rate distribution has a peak structure 
 and the details of the peak are consistent with the classical survival probability 
time distribution. We also investigate the behavior of the complex resonance
positions at the small opening limit ($n\rightarrow \infty$,
$n$ is the refractive index of the cavity).
At the large $n$ limit, the real
part of complex resonance positions approaches the solutions
with different $m$
of Dirichlet problem with a scale $n^{-2}$ and the imaginary
part goes zero as $n^{-2m}$ for TM and $n^{-2(m+1)}$ for TE
polarization, where $m$ is the order of the resonance.

\end{abstract}
\maketitle
\narrowtext

Two-dimensional (2-D) billiards with various boundary geometries
have been widely studied in quantum chaos community because
of the ease in analysis and rich interesting phenomena such
as classical and wave chaos and wave localization etc.
There are many relevant experimental realizations, e.g.,
surface waves \cite{Blu92}, microwave billiards \cite{Chi96,Chi97},
mesoscopic structures \cite{Sto90,Haa91,Sto99}, and semiconductor
microcavities \cite{Cha96,Gma98,Har03,Tan07,Flo06}.
In experiments, the systems measured are inevitably coupled
with environment, i.e., they are open.
Even in the case of a small coupling, it is sometimes difficult
to explain the open system based on the physical properties of
the corresponding closed system without a clear understanding of
the correspondence between them.

The decay-rate statistics is determined by interpaly 
of openness and dynamics of the system concerned\cite{Bor91}.
In practical purpose, it is, however, not a simple task to obtain 
many decay rates enough to apply statistical processes for a
chaotic open quantum system, such as a chaotic dielectric cavity,
because of heavy numerical tasks.
From the viewpoint of a easy analysis,
the circular dielectric cavity is a good open system, due
to its simple geometry, to study the statistics of decay rates
which originate from the dielectric property and the
relationship to the corresponding closed billiard with the
Dirichlet boundary condition.
In dielectric cavities the degree of opening
is determined by the refractive index $n$ and the small opening
limit corresponds to $n\rightarrow \infty $.

In this letter, we answer the following two questions
about resonances in a circular dielectric cavity:
(i) {\it How are the imaginary values of resonance positions,
which represent the decay rates of the resonances,
in the circular dielectric cavity distributed?}
(ii) {\it Would its resonance positions approach the eigenvalues
of the corresponding billiard at the small opening limit?}
For the first question, we obtain the complex resonance positions and
show that the distribution of imaginary values is consistent
with survival probability time distribution (SPTD) which represents
the decay property due to classical ray escapes \cite{Ryu06}.
As for the second question, we show that the resonance of TM polarization
with angular quantum number $m$ in circular dielectric cavity 
approaches the eigenvalues with $m-1$, not $m$, of Dirichlet problem
at the small opening limit ($n\rightarrow \infty$).

From the mathematical viewpoint, the only difference between closed
billiard and open dielectric cavity problems is the boundary
condition applied to the Helmholtz equation \cite{Jac75},
\begin{equation}
\label{helm}
(\nabla^2+k^{2})\psi = 0,
\end{equation}
where $k$ is the wavenumber inside the billiard or the dielectric
cavity. In the closed billiard problem, the typical boundary
conditions are Dirichlet and Neumann boundary conditions,
\begin{eqnarray}
\label{match0}
&(\mathrm{Dirichlet})& \psi({\bf r}_b)=0,\\\nonumber
&(\mathrm{Neumann})& \partial_n \psi ({\bf r}_b)=0,
\end{eqnarray}
where ${\bf r}_b$ denotes the boundary vector and $\partial_n$ is the
normal derivative to the boundary. For an arbitrary boundary shape
we can find {\it real} eigenvalues $k_n$ and the eigenfunctions
$\psi_n ({\bf r})$ satisfying the corresponding boundary condition.
In the dielectric cavity case, we have to apply different boundary
conditions depending on the polarization. For TM (TE) polarization, the
electric (magnetic) field normal to the 2-D cavity and its normal
derivative (its normal derivative divided $n^2$) are continuous at
the boundary interface, i.e.,
\begin{eqnarray}
\label{match1}
&(\mathrm{TM})& \frac{\psi_{in}({\bf r}_b)}{\partial_n \psi_{in}({\bf r}_b)}
=\frac{\psi_{out}({\bf r}_b)}{\partial_n \psi_{out}({\bf r}_b)},\\\nonumber
&(\mathrm{TE})& \frac{n_{in}^2\psi_{in}({\bf r}_b)}{\partial_n \psi_{in}({\bf r}_b)}
=\frac{n_{out}^2\psi_{out}({\bf r}_b)}{\partial_n \psi_{out}({\bf r}_b)},
\end{eqnarray}
where $\psi_{in}({\bf r})$ and $\psi_{out}({\bf r})$ are wave functions
inside and outside the dielectric cavity, respectively, and the refractive
indices of the cavity and environment are $n_{in}$ and $n_{out}$.
We will set $n_{in}=n$ and $n_{out}=1$ throughout the letter.
The resonance positions $k$, obtained from the above boundary conditions,
are complex numbers with negative imaginary values. The quality(Q)
factor of the corresponding resonance mode is defined as
$-\mbox{Re}(k)/2\mbox{Im}(k)$. High Q resonance modes, therefore,
have smaller absolute values of $\mbox{Im}(k)$ and mean the resonance
modes well confined inside the cavity.

For the circular boundary geometry with a radius $R$, the Helmholtz equation,
Eq.(\ref{helm}), can separate into the angular and radial equations due to
the rotational symmetry, and the solution of the angular equation is
simply given by $e^{im \phi}$ ($m$ is an integer).
The radial equation is represented by Bessel differential equation \cite{Gra00},
and relevant solution inside both the circular billiard and dielectric cavity is
Bessel function, $J_m(kr)$. In the circular dielectric cavity, the solution outside
the cavity should be outgoing wave which is represented by Hankel function of
the first kind, $H^{(1)}_m(kr)$. Using these solutions, the boundary conditions
(Eq.(\ref{match0}), Eq.(\ref{match1})) become \cite{Jac75, Noc97, Hen02}
\begin{eqnarray}
\label{match3}
&(\mathrm{Dirichlet})&J_{m}(k)=0,\\\nonumber
&(\mathrm{Neumann})&kJ_{m-1}(k)-mJ_{m}(k)=0,\\\nonumber
&(\mathrm{TM})&nJ_{m-1}(k)H_{m}^{(1)}(k_{0})-J_{m}(k)H_{m-1}^{(1)}(k_{0})=0,\\\nonumber
&(\mathrm{TE})&nJ_{m}(k)H_{m-1}^{(1)}(k_{0})-J_{m-1}(k)H_{m}^{(1)}(k_{0})\\\nonumber
&&=\frac{m}{k_{0}}(n-\frac{1}{n})J_{m}(k)H_{m}^{(1)}(k_{0}),
\end{eqnarray}
where we take $R=1$ without loss of generality, and $k_{0}$ is the vacuum wavenumber
and $k_{0}=k/n$. In general, the separability of the Helmholtz equation implies
that the circular system is integrable and there are good quantum numbers
specifying the solution. In our case, the solution can be classified by
the angular quantum number $m$ and the radial quantum number $l$ as $k_{(m,l)}$.
Figure 1 shows eigenfunctions satisfying Dirichlet (a) and Neumann (b)
boundary conditions and resonance modes in TM (c) and TE (d) cases, and these are
specified by a mode index $(\pm m,l)=(\pm 8,1)$. The eigenvalues and the resonance
positions are $k_{D(\pm 8,1)}=12.2251$ (a), $k_{N(\pm 8,1)}=9.6474$ (b),
$k_{TM (\pm 8,1)}=10.7845-i0.02278$ (c), and $k_{TE (\pm 8,1)}=11.6295-i0.03993$,
where we take
$n=2$ in TM and TE cases. It is easily seen that the numbers of high intensity
spots along the perimeter (angular direction) and along the radial direction
are $2m$ and $l$, respectively. As expected from the above example, the eigenvalues
and the resonance positions for one mode index $(\pm m,l)$ have an order as
$k_{D(\pm m,l)} > \mbox{Re}[k_{TE(\pm m,l)}] > \mbox{Re}[k_{TM(\pm m,l)}]
 > k_{N(\pm m,l)}$.

We confirm numerically that all resonances in the TM case can be classified
by the mode index $(\pm m,l)$ just like the Dirichlet and Neumann cases.
However, we find that in the TE case there are additional modes with
a nonzero angular momentum $m$, absent in other cases, which can not be
classified by the mode index $(\pm m, l)$.
Figure 2 shows an example of the additional modes, $(\pm m,l)=(\pm 8,\times)$
and $k_{TE(\pm 8,\times)}=17.3507-i2.4802$. As expected from the figure and large
absolute value of $\mbox{Im}(k)$, these modes are very leaky and
originated from the existence of the Brewster angle in TE case
on which rays can transmit without reflection. Therefore, these modes
do not show any notable wave confinement by the dielectric interface.

In order to investigate the distribution of resonance positions of the
circular dielectric cavity, we obtain all solutions in the range of
$0<\mbox{Re}(kR)<150$ with $R=1$ and $n=2$ for both TM and TE cases.
With the real part of the complex solutions obtained, we can check
the level spacing distribution that is known to be Poisson and
Wigner distribution for the integrable and chaotic billiards, respectively \cite{Sto99}.
Although the circular dielectric cavity is an open system, it is still
integrable, equivalent with the fact that all resonances can be
specified by the mode index $(m,l)$. We can, therefore, expect that
the level spacing distribution is Poissonic, and as shown in Fig. 3
the numerical calculation confirms this expectation.

The properties of openness in the circular dielectric cavity have been
investigated by calculating the SPTD \cite{Ryu06}
which shows very different short time behaviors depending on the polarization.
Especially, the exponential short time behavior appears in TE case and it has
some relation to the existence of the Brewster angle.
It is natural to relate this ray dynamical result with the distribution
of imaginary values of the resonance positions since the imaginary value
is relevant to the wave confinement by the dielectric cavity.

The distributions of imaginary values of resonance positions for both TM and TE cases
are shown in Fig. 4 (a) and (b), respectively. The substantial difference between
both cases is that the distribution for the TM case is bounded, i.e., has a minimum
imaginary value $-\gamma_M(n)/2$ while the result for the TE case is unbounded
\cite{Hen02b}. This result can be easily understood
from details of reflection coefficients
$R_{TM}(n,\theta)$ and $R_{TE}(n,\theta)$, $\theta$ is the incident angle,
determined by the Fresnel equations \cite{Haw95}. Consider an initial ray with
a fixed incident angle $\theta$ which is invariant in the circular boundary.
The survival probability of the ray would decay as $I(t)=e^{-\gamma(n,\theta) t}$ where
the decay rate is given by
\begin{equation}
\label{gammacl}
\gamma(n,\theta)= -\frac{\ln R_{TM(TE)}(n,\theta)}{2 \cos \theta},
\end{equation}
and the time $t$ is scaled as the length of the ray trajectory.
On the other hand, the intensity of a resonance decays as
\begin{equation}
I(t)=| e^{-i\omega t}|^2=|e^{-i(\mbox{Re}(k)+i\mbox{Im}(k))t}|^2
=e^{2\mbox{Im}(k)t}.
\end{equation}
Therefore, the decay rate of a resonance mode $(m,l)$ is given by
\begin{equation}
\label{gammamode}
\gamma_{(m,l)}=-2\mbox{Im}(k_{(m,l)}).
\end{equation}
The ray dynamical (Eq.(\ref{gammacl})) and resonance mode (Eq.(\ref{gammamode}))
decay rates have the same physical meaning due to the invariance of incident angle
$\theta$ in the circular boundary case.
Therefore, if $\gamma (n,\theta)$ in Eq.(\ref{gammacl}) is bounded, we can say that
$\mbox{Im}(k_{(m,l)})$ is also bounded in TM case. We note that $R_{TM}(n,\theta)$
has its minimum at $\theta=0$, i.e., the case of normal incident rays or
bouncing ball trajectories. Then, the minimum $\mbox{Im}(k_{TM})$ is
$-\gamma_M(n)/2$ and
\begin{equation}
\gamma_M(n)=-\frac{\ln R_{TM}(n,\theta=0)}{2}=\ln \frac{n+1}{n-1}.
\label{gammaM}
\end{equation}
The minimum imaginary value for the $n=2$ case is $-(\ln 3)/2\simeq -0.55$ which
is consistent with Fig.4 (a). The above expression for the minimum imaginary
value shows very good agreement with numerical results for various refractive
indices, $n=2,4,6,8,10$, as shown in Fig 5 (a).
On the other hand, in TE case there is the Brewster angle $\theta_B=\arctan(1/n)$,
and the ray incident with $\theta_B$ transmits without reflection, i.e.,
$R_{TE}(n,\theta_B)=0$. Then the decay rate $\gamma (n,\theta)$ in Eq.(\ref{gammacl})
becomes infinity at $\theta=\theta_B$, and the imaginary value of resonance position
in TE case can have very large absolute value of $\mbox{Im}(k)$ like the additional
mode $(\pm m,\times)$ mentioned before. This fact is consistent with the numerical
result in Fig. 4 (b).

We emphasize that the distributions of imaginary values of resonance positions,
excepting the very high-Q resonances with only tunneling decay,
can explain the behavior of the SPTD \cite{Ryu06}.
The resonance modes near the peak in Fig. 4 contribute to the short-time
exponential decay in the SPTD, and the high-Q modes distributed just below
the tunneling regime near zero in Fig. 4 do to the long-time
algebraic behavior in the SPTD. In TM case, the fact that two mode-classes are
smoothly connected, is consistent with the gradual
transition from exponential to algebraic decays. On the other hand,
in TE case, the distribution show an abrupt change at the peak, i.e., the
two mode-classes are almost isolated, and this is responsible to
the clear transition point from exponential to algebraic decays in SPTD.

In order to understand whole shape of the distributions, we plot the resonance
positions $k_{(m,l)}$ with $m=40$ in Fig. 5 (b)\cite{Hen02,Hen02b}. 
Black rectangle and red diamond
represent resonance positions of TM and TE cases, respectively.
It is known that the incident angle $\theta$ of waves in a resonance mode
$(m,l)$ can be estimated by a semiclassical relation \cite{Cha96}
\begin{equation}
\label{eikonal}
\sin{\theta} = m/\mathrm{Re}(k_{(m,l)}).
\end{equation}
From this relation we can obtain some $\mbox{Re}(k)$ values corresponding
to the critical angle $\theta_c$ for total internal reflection and
the Brewster angle $\theta_B$, which are indicated by black and red arrows,
respectively, in Fig. 5 (b). If the incident angles of resonance modes calculated
from Eq. (\ref{eikonal}) are greater than the critical angle $\theta_c$,
waves inside are very well confined by total internal reflection and
then the resonance mode has very small $|\mbox{Im}(k)|$ as shown on the left of
the black arrow. These high Q modes
explain the peak near $|\mbox{Im}(k)|=0$ in Fig. 4 (a) and (b), and
the fraction of these high Q modes would be $(1-1/n)$ in the semiclassical
limit. As $\mbox{Re}(k)$ increases beyond the black arrow, $\mbox{Im}(k)$
converges gradually to the minimum value $-\gamma_M/2$ in the TM case, which
corresponds to the bounded distribution near the minimum value in Fig. 4 (a).
In the TE case, near the red arrow corresponding to $\theta_B$,
$\mbox{Im}(k)$ shows very low value and then converges to $-\gamma_M/2$,
which explains the unbounded distribution and the peak at $-\gamma_M/2$ in Fig. 4 (b).
The same limit value $-\gamma_M/2$ can be expected from the fact that
$R_{TM}(n,\theta=0)=R_{TE}(n,\theta=0)$.
The green and blue lines are obtained by calculating the relation between
$\mbox{Re}(k)$ and $\mbox{Im}(k)$ from Eq. (\ref{gammacl}), (\ref{gammamode}),
and (\ref{eikonal}). These explain well the resonance positions on the right
range of the black arrow in Fig. 5 (b).

Under the assumption that $\sin \theta$ values of resonance
modes distribute uniformly such that  $P(\sin \theta)=constant$, we can
obtain the distribution
of $\mbox{Im}(k)$ of resonance modes as
\begin{equation}
\label{prob}
P(\mbox{Im} (k))\propto  ( \frac{d (\mbox{Im}(k))}{d(\sin \theta)} ) ^{-1}.
\end{equation}
Using Eq. (\ref{gammacl}) and (\ref{gammamode}), we can calculate
the distributions for both TM and TE cases, and the results are
the red lines in Fig. 4 (a) and (b) which are in a good agreement
with the histograms near the minimum imaginary value.

Now we discuss the behavior of resonance positions at the small
opening limit, i.e., $n \rightarrow \infty$. Since the imaginary
value of a resonance position $k_{(\pm m,l)}$ means the decay
rate as shown in Eq. (\ref{gammamode}), we can safely take
the limiting resonance position as a {\it real} value $k_L$.
We first focus on the TM case. When $n$ is very large, the solution of
the boundary condition for TM polarization shown in Eq.(\ref{match3})
would be $k=k_L+\delta$ where $\delta$ would be a small complex number
and be zero at $n \rightarrow \infty$. We can then expand
the boundary equation Eq.(\ref{match3}) around $k_L$.
The Bessel functions can be written as
\begin{eqnarray}
\label{expand_JH}
J_{m}(k_L +\delta) &\simeq& J_{m}(k_L)+ J_{m}'(k_L)\cdot \delta +\cdots,\\\nonumber
   &=&  J_{m}(k_L)+[-J_{m+1}(k_L)+\frac{m}{k_L} J_m(k_L)]\cdot \delta+\cdots,
\end{eqnarray}
where $J_{m}'(k_L)=\frac{dJ_{m}(k)}{dk}{\Big\vert}_{k=k_L}$, and
the ratio of Hankel functions becomes
\begin{equation}
\frac{H_{m-1}(k_L/n)}{H_m (k_L/n)} \simeq A n^{-1}+i B n^{-2m+1}
\end{equation}
by the approximation by tangents of Bessel functions \cite{Gra00,Noc97}.
The coefficients $A$ an $B$ is given by
\begin{eqnarray}
A&=&\frac{ek_L}{2(m-1)} (\frac{m-1}{m})^{m-1/2}, \\
B&=&\frac{1}{2}(\frac{ek_L}{2})^{2m-1}(\frac{1}{m(m-1)})^{m-1/2}.
\end{eqnarray}
Then, the expanded equation of the boundary condition has
small quantities, $1/n$ and $\delta$, and in the lowest order
($1/n=0$ and $\delta=0$), the equation becomes
\begin{equation}
J_{m-1}(k_L)=0.
\end{equation}
This is nothing but the Dirichlet boundary condition
(see Eq.(\ref{match3})), which means that at the small opening
limit the resonance positions with order $m$ of TM case approach
the eigenvalues with order $m-1$ of Dirichlet problem, i.e.,
\begin{equation}
\lim_{n\rightarrow \infty} k_{TM(\pm m,l)} = k_{D(\pm (m-1),l)}.
\end{equation}
From the equation for the next order, we know how the resonance
positions converge to the limiting value $k_c$. The result is
\begin{equation}
\delta_{TM} = -An^{-2}-iB n^{-2m}.
\label{deltatm}
\end{equation}
Note that the imaginary part has an exponent depending on
the angular quantum number $m$, while the exponent of the real part
does not depend on the mode indices.

For TE polarization, we can obtain similar results. In the lowest order,
the TE boundary condition becomes
\begin{equation}
J_m(k_L)=0.
\end{equation}
This implies
\begin{equation}
\lim_{n\rightarrow \infty} k_{TE(\pm m,l)} = k_{D(\pm m,l)}.
\end{equation}
Note that at the small opening limit the resonance positions
with order $m$ of TE case approach the eigenvalues with the {\it same}
order $m$ of Dirichlet problem.
From the equation for the next order, we get
\begin{equation}
\delta_{TE}=-\frac{k_L}{m}n^{-2} - i (\frac{k_L}{m})^2 B n^{-2(m+1)}.
\label{deltate}
\end{equation}
From the comparison between $\mbox{Im}(\delta_{TM})$ and
$\mbox{Im} (\delta_{TE})$, it is clear that the TE boundary condition
is more effective than the TM boundary condition in confining waves
inside the circular cavity when the mode incident angle are greater
than $\theta_c$.

In order to numerically confirm the above results,
we trace two resonances of mode indices $(m,l)=(8,1)$ and $(8,40)$
with increasing $n$ for both TM and TE cases. As expected from the
analytical results, we obtain
\begin{eqnarray}
\lim_{n\rightarrow \infty} k_{TM(8,1)}&=& k_{D(7,1)}\simeq 11.0864,  \\
\lim_{n\rightarrow \infty} k_{TM(8,40)} &=& k_{D(7,40)} \simeq 135.6942, \\
\lim_{n\rightarrow \infty} k_{TE(8,1)} &=& k_{D(8,1)} \simeq 12.2251, \\
\lim_{n\rightarrow \infty} k_{TE(8,40)} &=& k_{D(8,40)} \simeq 137.2123.
\end{eqnarray}
In Fig. 6 and Fig. 7, we show the $n$-dependence of $\delta_{TM}$ and $\delta_{TE}$.
The results of Eq.(\ref{deltatm}) and (\ref{deltate}), denoted by solid lines,
explains well the small opening limit behavior of resonance positions.
It is noted that there is a transition point at $n=n_c\simeq 17$ in the traces of
both $k_{TM(8,40)}$ and $k_{TE(8,40)}$ where the incident angle of the
resonance modes(see Eq.(\ref{eikonal})) becomes the critical angle
$\theta_c$ for total internal reflection. Therefore, the transition
point indicates the change of leakage mechanism, from refractive
to tunneling leakage of waves. For the refractive leakage range ($n<n_c$),
the trace of TM resonance modes shows $1/n$ dependence \cite{Noc97}.

In summary, we have studied the distribution of resonance positions $k_{(m,l)}$
in the circular dielectric cavity, and the behavior of  $k_{(m,l)}$
at the small opening limit for both TM and TE polarizations.
The distributions of $\mbox{Im}(k_{TM(m,l)})$
and  $\mbox{Im}(k_{TE(m,l)})$ are consistent with corresponding 
ray dynamical SPTD behaviors, and the small opening limits ($n\rightarrow \infty$)
of  $k_{TM(m,l)}$ and $k_{TE(m,l)}$  approach
different eigenvalues of Dirichlet problem, $k_{D(m-1,l)}$ and
$k_{D(m,l)}$, respectively.
The result implies that the small opening limit of some open system
does not directly match to the corresponding closed system as shown
in the TM case.

\begin{center}
{\bf Acknowledgments}
\end{center}
This work was supported by the Creative Research Initiatives (Center for
Quantum Chaos Application) MOST/KOSEF. J.-W. R. and S.-Y. L. were
supported by the Brain Korea 21 Project in 2006, and C.-M. K. is partially
supported by Sogang Research Grant of 20071114.

\newpage
\begin{center}
{\bf Figure Captions}
\end{center}

\noindent {\bf FIG. 1:} The intensity patterns of a mode $(m,l)=(8,1)$
    for various boundary conditions; (a) Dirichlet,
    (b) Neumann, (c) TM, and (d) TE boundary conditions.

\noindent {\bf FIG. 2:} The intensity pattern of the additional resonance mode
$(\pm m,l)=(\pm 8,\times)$ in the TE case. This corresponds to a very leaky mode
showing wave escaping through the Brewster angle $\theta_B$.

\noindent {\bf FIG. 3:} The level spacing distributions. (a) TM case.
(b) TE case. These show good agreements with Poisson
distribution (the solid line).

\noindent {\bf FIG. 4:} The imaginary value distributions of resonance positions
when $n=2$; (a) TM case and (b) TE case.
Red lines are the results of a semiclassical analysis, Eq.(\ref{prob}).

\noindent {\bf FIG. 5:} (a) The minimum imaginary values of the distributions 
for $n=2,4,6,8,10$ in TM case are denoted by solid dots. The solid line 
represents $-\gamma_M (n,\theta=0)/2$ and $\gamma_M(n,\theta=0)$
(see Eq. (\ref{gammaM})) is the decay rate of bouncing ball trajectory.
(b) The resonance positions with a fixed angular quantum number, $m=40$,
and $n=2$ for both TM (rectangles)   and TE (diamonds) cases. 
The left and right arrows represent $\mbox{Re}(k)$ values where the incident
angle of resonance modes (see Eq. (\ref{eikonal})) becomes
to the critical angle $\theta_c$ and the Brewster angle $\theta_B$, respectively.
The resonance positions with incident angle greater than $\theta_c$ are well 
explained by results of the semiclassical analysis described by 
green (TM) and blue (TE) solid lines.
Red cross represents an additional resonance mode which is originated from
the existence of the Brewster angle.

\noindent {\bf FIG. 6:} The approaching behaviors of real part
of the resonance positions corresponding to mode indices $(m,l)=(8,1)$
 and $(8,40)$ with increasing $n$ (the small opening limit). 
 The solid lines are the real parts (first term) of Eq.(\ref{deltatm}) and
(\ref{deltate}) showing $n^{-2}$ behavior. 
\noindent {\bf FIG. 7:} The approaching behaviors of imaginary part
of the resonance positions corresponding to mode indices $(m,l)=(8,1)$
 and $(8,40)$ with increasing $n$ (the small opening limit).
 The solid lines are the imaginary part (second term) of Eq.(\ref{deltatm}) 
 and (\ref{deltate}) showing $n^{-16}$ (TM) and $n^{-18}$ (TE).

\newpage

\begin{figure}
\begin{center}
\includegraphics[width=0.45\textwidth]{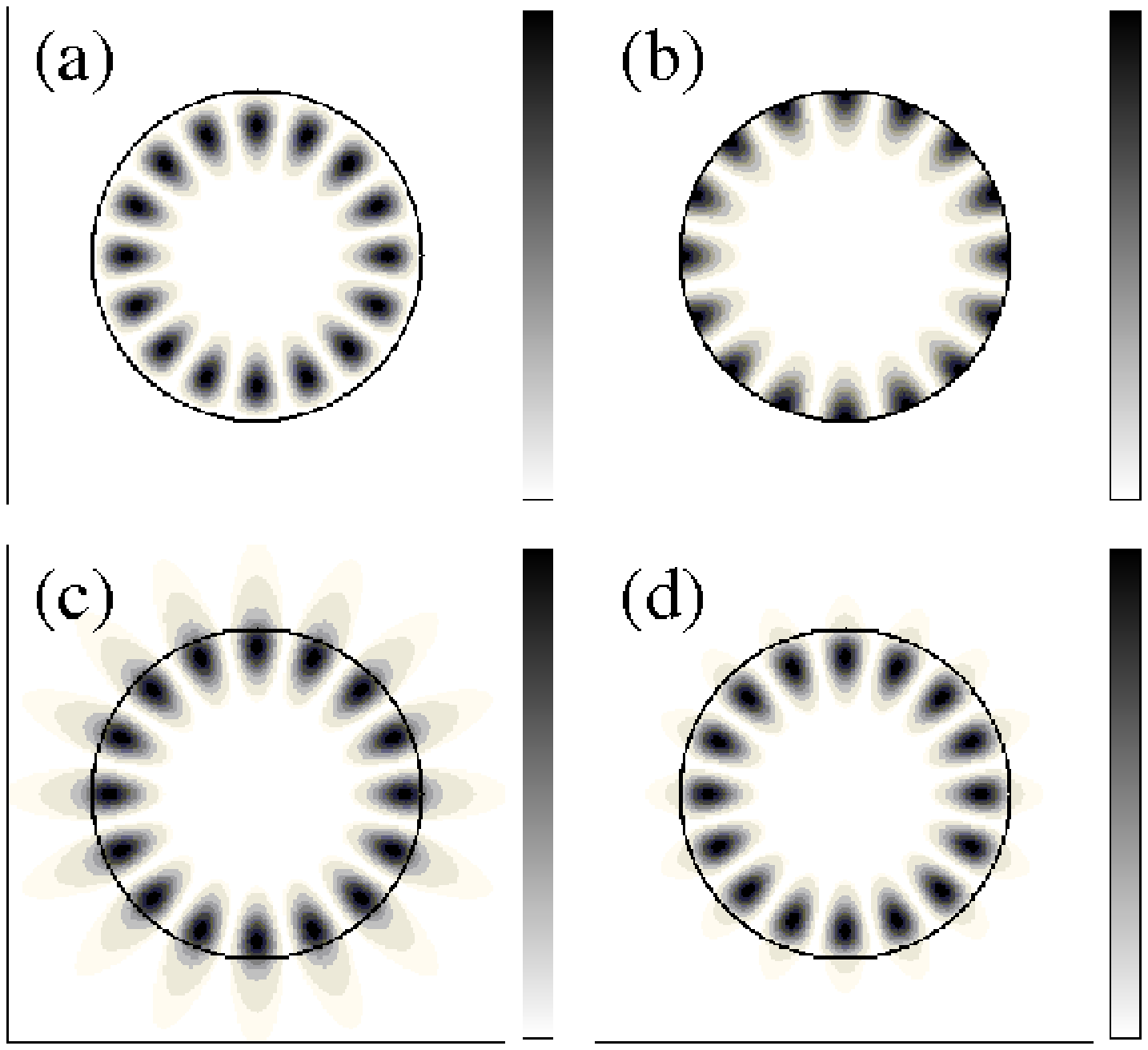}
\caption{}
\end{center}
\end{figure}
\newpage

\begin{figure}
\begin{center}
\includegraphics[width=0.3\textwidth]{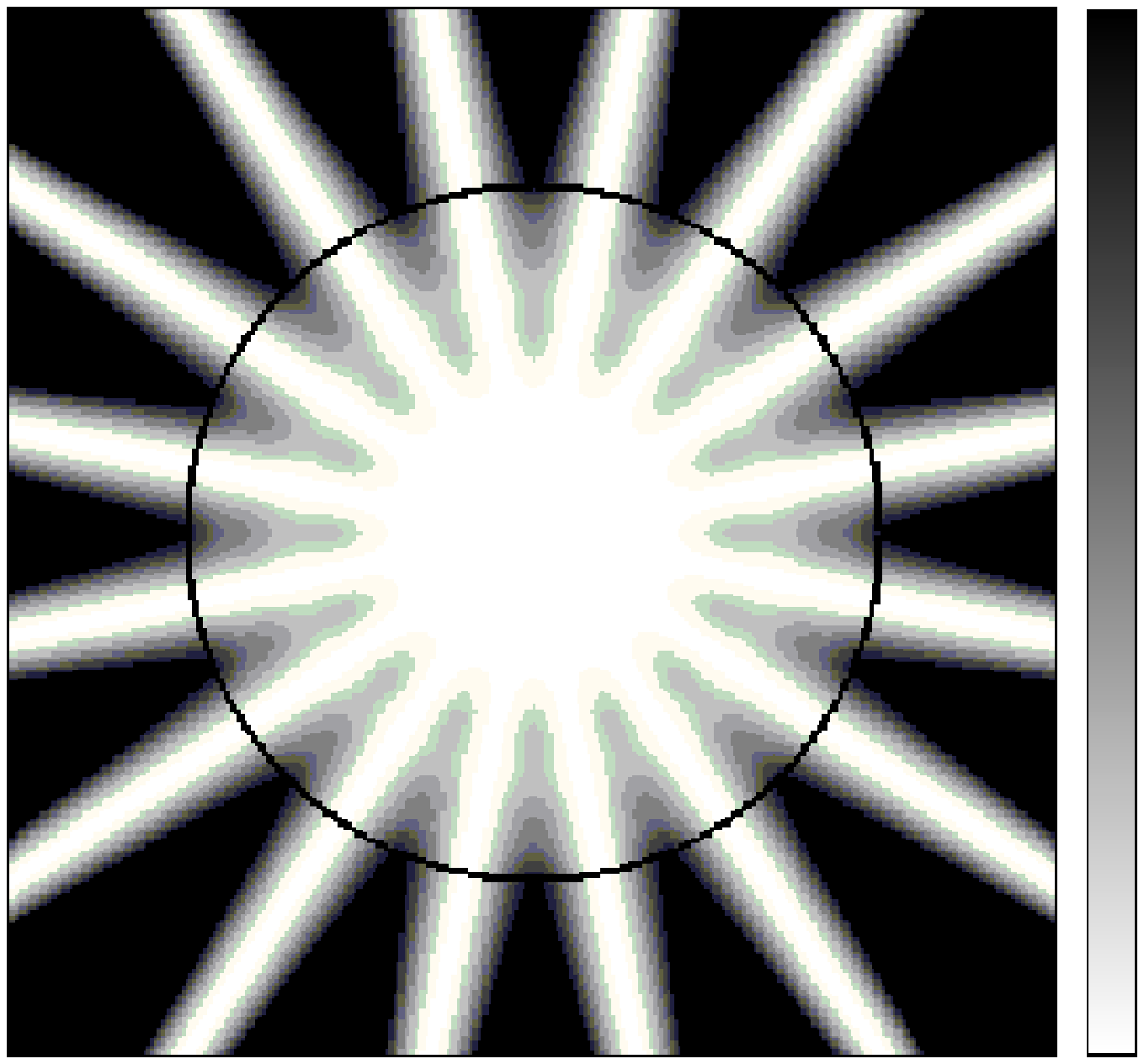}
\caption{}
\end{center}
\end{figure}
\newpage

\begin{figure}
\begin{center}
\includegraphics[width=0.4\textwidth]{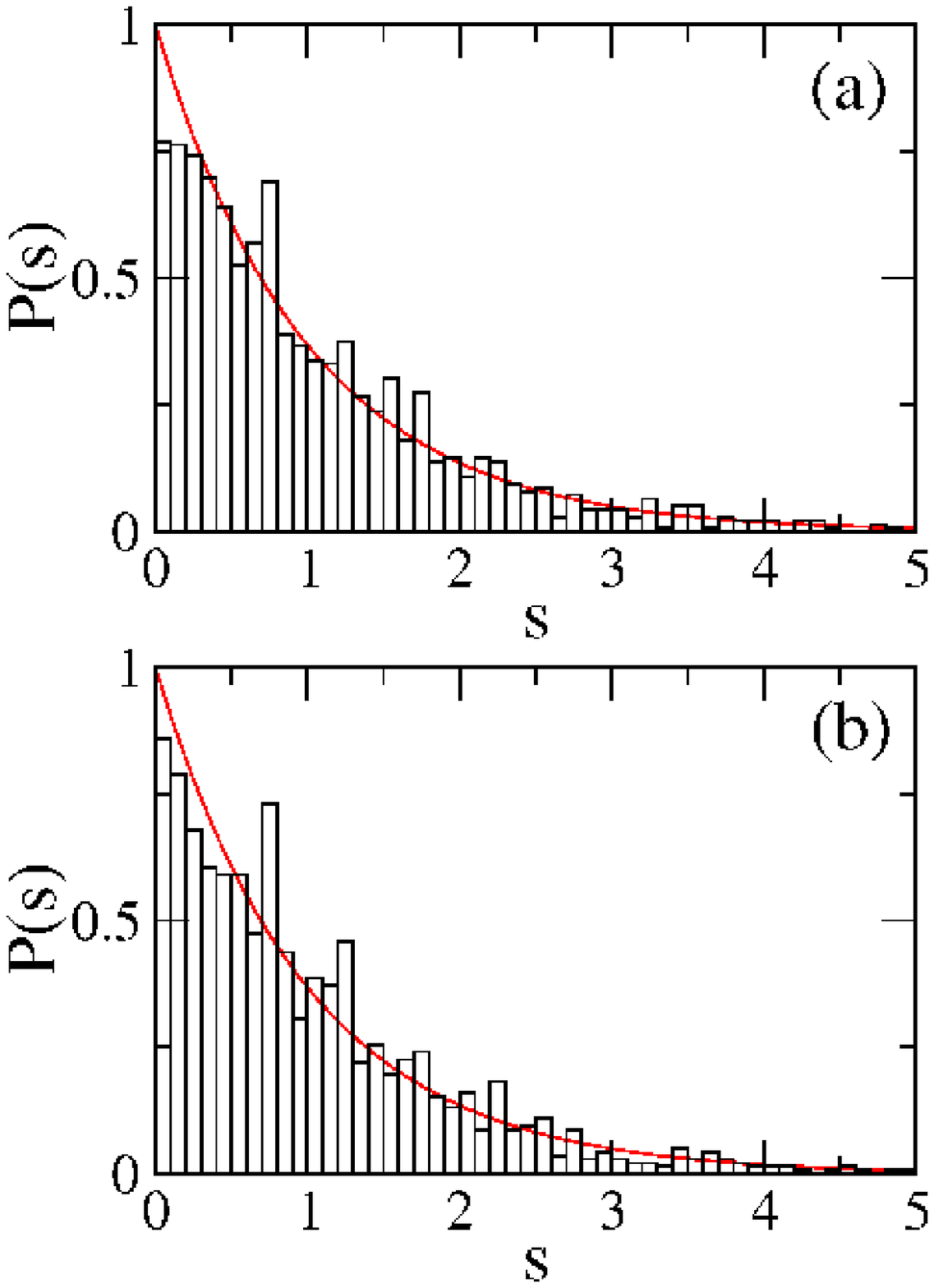}
\caption{}
\end{center}
\end{figure}
\newpage

\begin{figure}
\begin{center}
\includegraphics[width=0.4\textwidth]{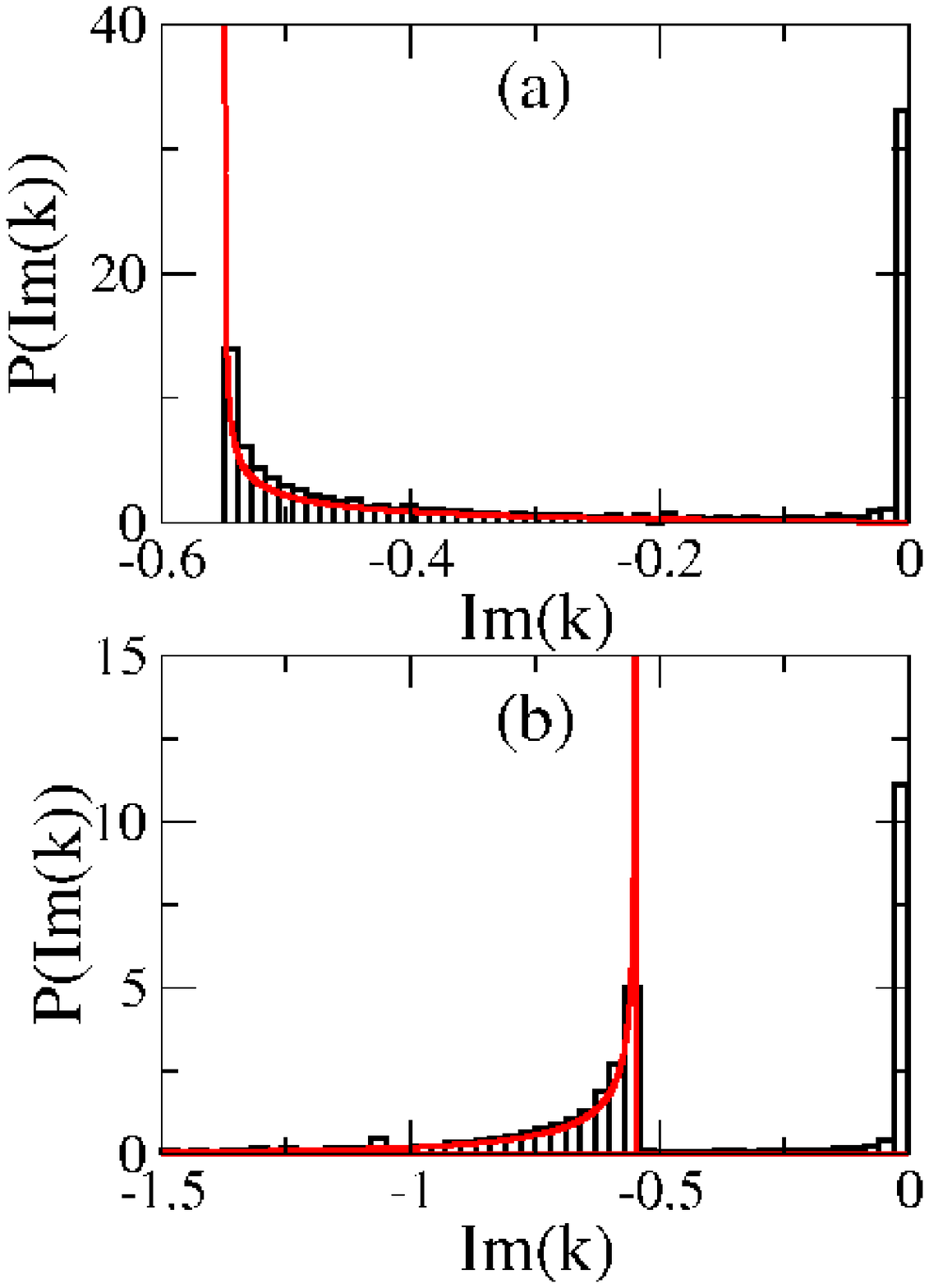}
\caption{}
\end{center}
\end{figure}
\newpage

\begin{figure}
\begin{center}
\includegraphics[width=0.4\textwidth]{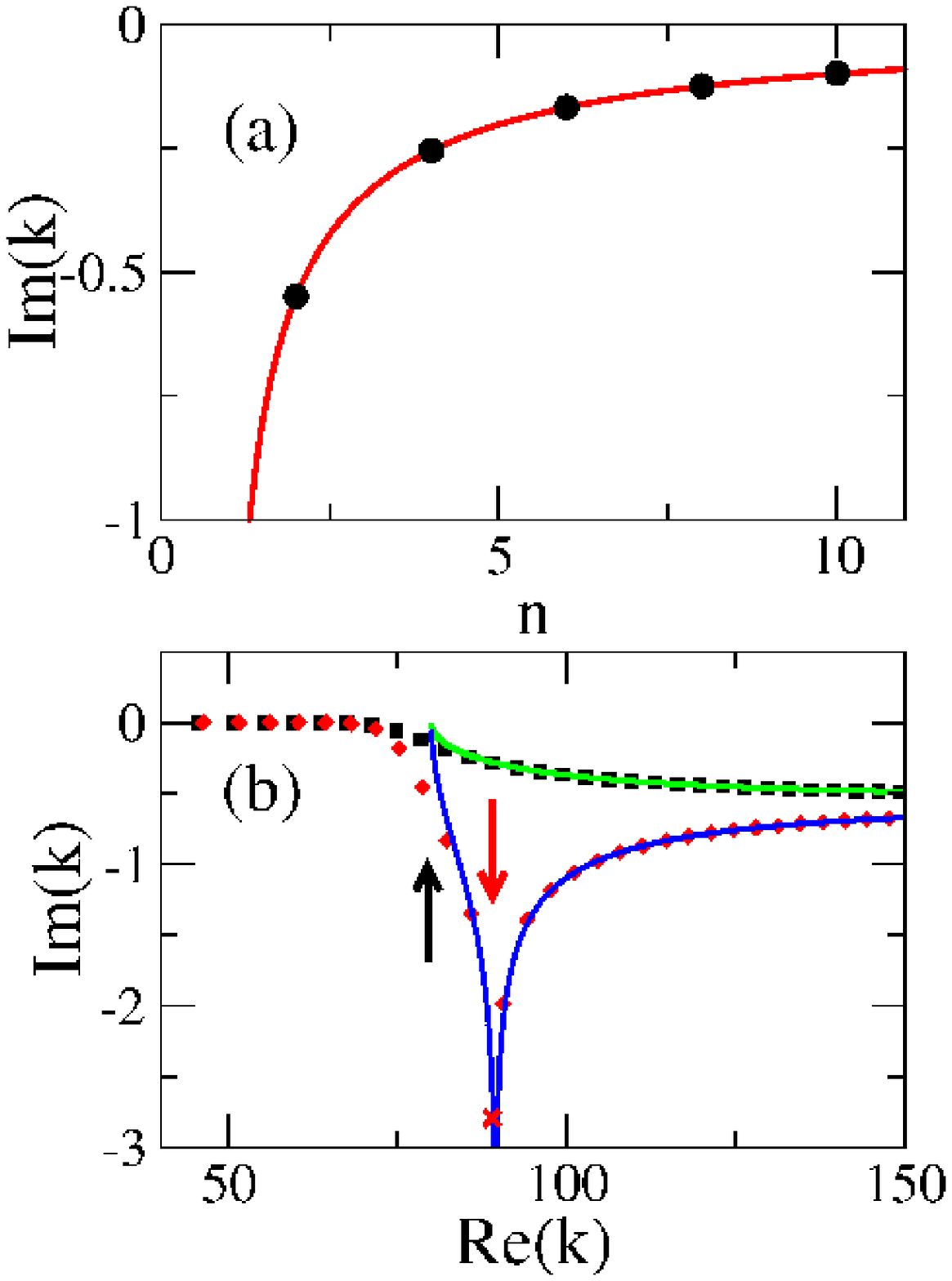}
\caption{}
\end{center}
\end{figure}
\newpage

\begin{figure}
\begin{center}
\includegraphics[width=0.45\textwidth]{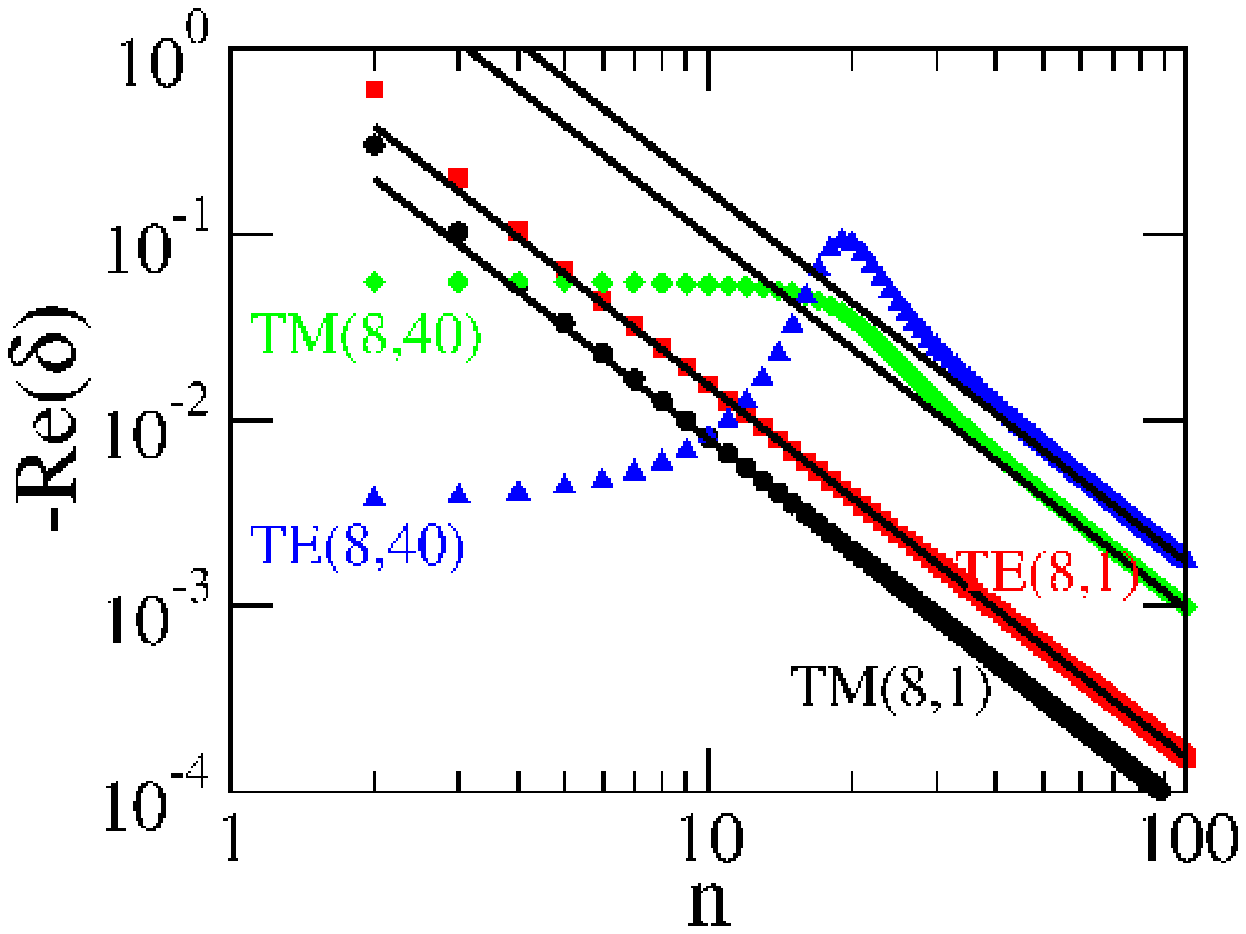}
\caption{}
\end{center}
\end{figure}
\newpage

\begin{figure}
\begin{center}
\includegraphics[width=0.45\textwidth]{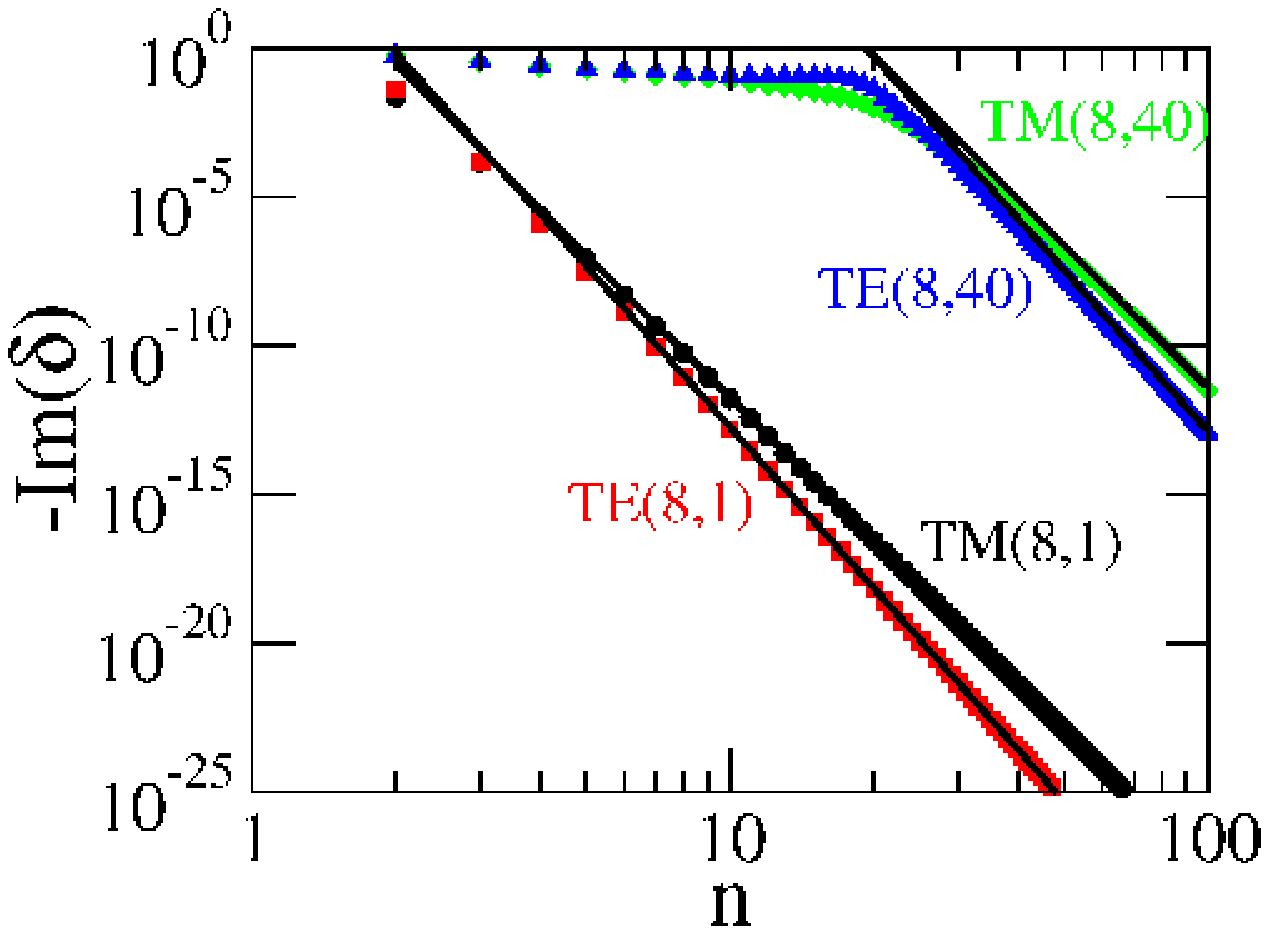}
\caption{}
\end{center}
\end{figure}

\begin{thebibliography}{150}

\bibitem{Blu92} R. Bl\"umel, I. H. Davidson, W. P. Reinhardt, H. Lin, and
M. Sharnoff, Phys. Rev. A 45 (1992) 2641.
\bibitem{Chi96} P. A. Chinnery and V. F. Humphrey, Phys. Rev. E 53 
(1996) 272.
\bibitem{Chi97} P. A. Chinnery, V. F. Humphrey, and C. Beckett,
J. Acoust. Soc. Am. 101 (1997) 250.
\bibitem{Sto90} H. -J. St\"ockmann and J. Stein, Phys. Rev. Lett. 64
(1990) 2215.
\bibitem{Haa91} F. Haake, G. Lenz, P. Seba, J. Stein, H. -J. St\"ockmann,
and K. Zyczkowski, Phys. Rev. A 44 (1991) R6161.
\bibitem{Sto99} H.-J. St\"ockmann, {\it Quantum Chaos; An Introduction}
(Cambridge University Press, UK, 1999) references therein.
\bibitem{Cha96} {\it Optical Processes in Microcavities},
edited by R. K. Chang and A. J. Campillo (World Scientific, Singapore, 1996).
\bibitem{Gma98} C. Gmachl, F. Capasso, E. E. Narimanov, J. U. N\"ockel,
A. D. Stone, J. Faist, D. L. Sivco, and A. Y. Cho, Science 280
(1998) 1556.
\bibitem{Har03} T. Harayama, P. Davis, and K. S. Ikeda, Phys. Rev. Lett.
90 (2003) 063901.
\bibitem{Tan07} T. Tanaka, M. Hentschel, T. Fukushima, and T. Harayama,
Phys. Rev. Lett. 98 (2007) 033902.
\bibitem{Flo06} J.-M. le Floch, J. D. Anstie, M. E. Tobar, J. G. Hartnett, P.-Y. Bourgeois,
and D. Cros, Phys. Lett. A 359 (2006) 1.
\bibitem{Bor91} F. Borgonovi, I. Guarneri, and D. L. Shepelyansky,
Phys. Rev. A 43 (1991) 4517 .
\bibitem{Ryu06} J.-W. Ryu, S.-Y. Lee, C.-M. Kim, and Y.-J. Park,
Phys. Rev. E 73 (2006) 036207.
\bibitem{Jac75} J. D. Jackson, {\it Classical Electrodynamics} 2nd Edition
(John Wiley \& Sons, New York, 1975).
\bibitem{Gra00} I. S. Gradshteyn, and I. M. Ryzbik, \textit{Table of Integrals,
Series, and Products}, 6th Edition (Academic Press, San Diego, 2000).
\bibitem{Noc97} J. U. N\"ockel, Ph.D. thesis, Yale University, 1997.
\bibitem{Hen02} M. Hentschel, Ph.D. thesis,
Max Planck Institute for the Physics of Complex Systems, 2002.
\bibitem{Hen02b} M. Hentschel and J. U. N\"ockel, physics/0203064 (2002).
\bibitem{Haw95} J. Hawkes and I. Latimer, \textit{Lasers; Theory and Practice}
(Prentice Hall, Englewood Cliffs, NJ, 1995).

\end{thebibliography}
\end{document}